\documentclass[nofootinbib,floatfix,superscriptaddress,twocolumn]{revtex4}
\usepackage{graphicx}
\usepackage{epsfig}
\usepackage{bm}
\usepackage[T1]{fontenc}
\usepackage[latin9]{inputenc}
\usepackage{amssymb}
\usepackage{float}
\usepackage{amsmath}
\usepackage{dcolumn}
\usepackage{cancel}
\usepackage[colorlinks]{hyperref}
\usepackage[usenames,dvipsnames]{color}
\hypersetup{
     breaklinks=true,
    pdfstartview={FitH},    
    colorlinks=true,       
    linkcolor=blue,          
    citecolor=red,        
    filecolor=magenta,      
    urlcolor=blue,           
    anchorcolor=green,      
    linktocpage=true
}

\newcommand{\be}{\begin{equation}}
\newcommand{\ee}{\end{equation}}

\begin{document}

\title{Slow-roll inflation and growth of perturbations in Kaniadakis modification of Friedmann cosmology}

\author{G.~Lambiase}
\email{lambiase@sa.infn.it}
\affiliation{Dipartimento di Fisica, Universit\`a degli Studi di Salerno, Via Giovanni Paolo II, 132 I-84084 Fisciano (SA), Italy}
\affiliation{INFN Gruppo Collegato di Salerno - Sezione di Napoli c/o Dipartimento di Fisica, Universit\`a di Salerno, Italy}

\author{G.~G.~Luciano}
\email{giuseppegaetano.luciano@udl.cat}
\affiliation{Applied Physics Section of Environmental Science Department,  Escola Polit\`ecnica Superior, Universitat de Lleida, Av. Jaume
II, 69, 25001 Lleida, Spain}

\author{A. Sheykhi}
\email{asheykhi@shirazu.ac.ir}
\affiliation{Department of Physics, College of Sciences, Shiraz University, Shiraz 71454, Iran}
\affiliation{Biruni Observatory, College of Sciences, Shiraz University, Shiraz 71454, Iran}

\date{\today}

\begin{abstract}
Kaniadakis entropy is a one-parameter deformation of the classical Boltzmann-Gibbs-Shannon entropy, arising from a self-consistent relativistic statistical theory.  
Assuming a Kaniadakis-type generalization of the entropy associated with the apparent horizon of Friedmann-Robertson-Walker (FRW) Universe and using the gravity-thermodynamics conjecture, a new cosmological scenario is obtained based on the modified Friedmann equations. By employing such modified equations, 
we analyze the slow-roll inflation, driven by a scalar field with power-law potential, at the early stages of the Universe. 
We explore the phenomenological consistency of this model by computation of the scalar spectral index and tensor-to-scalar ratio.
Comparison with the latest BICEP and Planck data allows us to constrain Kaniadakis parameter to $\kappa\lesssim\mathcal{O}(10^{-12}\div10^{-11})$, which is discussed in relation to other observational bounds in the literature. We also disclose the effects of Kaniadakis correction term on the growth of perturbations at the early stages of the Universe by employing the spherically symmetric collapse
formalism in the linear regime of density perturbations. We find out that the profile of density contrast is non-trivially affected in this scenario. Interestingly enough, we observe 
that increasing Kaniadakis parameter $\kappa$ corresponds to a faster growth of perturbations in a Universe governed by the corrected Friedmann equations. Finally, we comment on the consistency of the primordial power spectrum for scalar perturbations with the best data-fit provided by Planck. 
\end{abstract}

 \maketitle

\section{Introduction}
\label{Intro} 

It is a general belief that our Universe has experienced two phases of accelerated expansion. The first, named inflation,  was proposed to address some internal problems (flatness, horizon, structure formation) of standard modern Cosmology~\cite{Guth,Linde,Staro}. 
The second is thought to have started roughly five billion years ago and has by now been confirmed by measurements of the luminosity distances of type Ia Supernovae~\cite{meas1,meas2}. 
While being largely understood in most of their facets, concerns remain about the origin of these phenomena.  
Among the candidate mechanisms, two explanations are the most credited so far: on one hand, it is possible to maintain the classical Einstein's description of gravity and introduce extra energy degrees of freedom that drive the Universe acceleration, such as an inflaton field~\cite{Bartolo:2004if} and the dark sectors of the cosmos~\cite{Peebles:2002gy,Cai:2009zp,Capolup}. 
Alternatively, one can resort to a class of modified theories of gravity and leave the energy content of the Universe unaffected (see~\cite{Capozziello:2011et} for a review). 
From the latter perspective, potential developments have been achieved in the gravity-thermodynamics picture, which conjectures that the gravitational field equations in the cosmological context can be extracted from the first law of thermodynamics on the Universe apparent horizon, and vice-versa~\cite{Jacobson:1995ab,Padmanabhan:2003gd,Padmanabhan:2009vy,Cai:2005ra,Akbar:2006kj,Cai:2006rs} (see~\cite{Paranjape:2006ca,Sheykhi:2007zp,Akbar:2006er,Sheykhi:2007gi,Jamil:2009eb,Sheykhi:2010zz,Fan:2014ala} for further applications).

In its traditional formulation, the gravity-thermodynamics conjecture identifies the thermodynamic entropy of the Universe with the Boltzmann-Gibbs-Shannon (BGS) entropy. In recent years, much efforts have been devoted to study generalized scenarios based on extended entropies, such as Tsallis~\cite{Tsallis1,Tsallis2,Tsallis3}, Barrow~\cite{Barrow:2020tzx} and Kaniadakis~\cite{Kania0,Kania1,Kania2,Scarf1,Scarf2} entropies, which all possess the BGS framework as a limit. In particular, Kaniadakis entropy is a one-parameter modification of the BGS entropy that arises from a coherent and self-consistent relativistic statistics, while still retaining the basic features of the classical BGS theory.
Motivated by the relativistic essence of this new formalism,
implications have been examined in several areas~\cite{Hasegawa,OSC,NC} and the problem of how to equip the Standard Model of Cosmology (SMC) with a fully relativistic description of the entropic degrees of freedom in the early Universe
has been investigated from multiple perspectives~\cite{Myrev}.

The modified Friedmann equations in cosmology based on Kaniadakis entropy have been first explored in~\cite{Lymperis:2021qty}, where the extra terms have been incorporated into additional dark energy components that reflect the effects of the corrected entropy. The ensuing theory has proved to exhibit a richer phenomenology compared to the SMC, especially as for the ability to predict the usual thermal history of the Universe~\cite{Lymperis:2021qty}, the evolution of the Hubble rate~\cite{Hernandez-Almada:2021aiw} and the early baryogenesis and primordial Lithium abundance~\cite{KanLuc}. Recently, an alternative derivation has been proposed in~\cite{Sheykhi:2023aqa} by incorporating Kaniadakis effects as geometrical (i.e. gravitational) corrections to the left-hand side of the Friedmann equations and keeping the energy content of the Universe as standard model. In this approach, the apparent horizon radius of the Universe varies with time due to the cosmic expansion \cite{Sheykhi:2023aqa}. 
In this context, it has been argued that the generalized second law of thermodynamics still holds for a Universe enclosed by the apparent horizon and endowed with Kaniadakis entropy. From theoretical shores, the analysis of Kaniadakis modification of Friedmann cosmology has landed on more experimental grounds in the last years. Observational attempts to constrain Kaniadakis corrections appear in~\cite{Hernandez-Almada:2021aiw,KanLuc,CosmKan4}, leading to different upper bounds on Kaniadakis parameter. 

Starting from the above premises, in this work we explore more in-depth the cosmological implications of Kaniadakis entropy. Specifically, we study the slow-roll inflationary model of the Universe driven by a scalar field with power-law potential. By computing the scalar spectral index and tensor-to-scalar ratio, we show that Kaniadakis modification of Friedmann cosmology is phenomenologically consistent with the latest BICEP and Planck data, provided that the entropic parameter is constrained to $\kappa\lesssim\mathcal{O}(10^{-12}\div10^{-11})$. Comparison with previous bounds on $\kappa$ supports the idea that the existing observational literature on Kaniadakis modification of Friedmann cosmology can be understood in a unified picture if one allows $\kappa$ to have a running behavior, namely to vary with the energy scale. We finally examine the influence of
Kaniadakis entropy corrections on the growth of cosmological perturbations in the early stages of the Universe. We employ the Top-Hat Spherical Collapse (SC) model~\cite{SCM}, which describes the evolution of uniform and spherical symmetric perturbations in an expanding
background by adopting the same Friedmann
equations for the underlying theory of gravity~\cite{Gr1,Gr2,Gr3}. We find out that the profile of density contrast
differs from the standard cosmology, in particular, the growth rate of the total perturbations becomes faster compared to the standard cosmology. Throughout this work we use the natural units 
$\hslash=c=k_B=1$.  

This paper is organized as follows. The next section is devoted to a review on the derivation of the modified Friedmann equations inspired by the Kaniadakis entropy. In section~\ref{Inf}, we 
explore the slow-roll solution for {inflation with a power-law potential} in entropy-corrected Kaniadakis modification of Friedmann cosmology. In section~\ref{Growth} we comment on the growth of perturbations at the early stages of the Universe using the spherically collapse formalism. We also also elaborate on the impact of Kaniadakis entropy on the primordial power spectrum for scalar perturbations.  Results are finally summarized in section~\ref{DC}. 

\section{Kaniadakis entropy and modified Friedmann equations}
\label{MFE}
Motivated by the evidence that the spectrum of the relativistic cosmic rays exhibits a power-law tailed behavior instead of the classical exponential one, the problem of deriving the generalization of the Maxwell-Boltzmann distribution in a special relativistic framework has been first addressed in~\cite{Kania1}. As a result, it has been shown that the symmetries of the Lorentz transformations impose the following modification of the BGS entropy 
\begin{equation}
\label{KE}
S_{\kappa}=-\sum_{i}n_i \ln_\kappa n_i\,,
\end{equation}
where 
\begin{equation}
 \label{logk}
\ln_{\kappa}x \equiv \frac{x^\kappa-x^{-\kappa}}{2\kappa}\,.
\end{equation}
Using the maximum entropy principle, the corresponding Boltzmann factor for the $i$-th level of a given system becomes 
\begin{equation}
n_i=\alpha \exp_\kappa\left[-\beta\left(E_i-\mu\right)\right],
\end{equation}
where 
\begin{equation}
\alpha=\left[(1-\kappa)/(1+\kappa)\right]^{1/2\kappa}\,,\quad\,\, 1/\beta=\sqrt{1-\kappa^2}\,\hspace{0.2mm}T\,,
\end{equation}
and the $\kappa$-deformed exponential is given by
\begin{equation}
\label{expk}
\exp_\kappa(x)\,\equiv\,\left(\sqrt{1+\kappa^2\,x^2}\,+\,\kappa\,x\right)^{1/\kappa}\,. 
\end{equation}

From Eq.~\eqref{KE}, it is evident that departure from the standard BGS entropy is quantified by the (dimensionless) exponent $-1<\kappa<1$. The classical framework is recovered in the limit of $\kappa\rightarrow0$.

Kaniadakis entropy is formally adopted for black holes to explore relativistic corrections to characteristic thermodynamic quantities, like the Hawking temperature, mass and heat capacity. These studies are also useful for holographic applications. In this context, it is convenient to express Eq.~\eqref{KE} in the probabilistic language~\cite{Jeans1,Jeans2}
\be
\label{KanP}
S_\kappa=-\sum_{i=1}^W \frac{P_i^{1+\kappa}-P_i^{1-\kappa}}{2\kappa}\,,
\ee
where $P_i$ is the probability for the system to be in the $i$-th microstate and $W$ the total number of permitted configurations. 
For equiprobable states (i.e. $P_i=1/W$) and recalling that the BGS entropy obeys $S\propto\log W$, we have $P_i=e^{-S}$. For the case of black holes ($S=S_{BH}=A/(4G)$), this takes the form $P_i=e^{-A/(4G)}$. Here, we have denoted the standard Bekenstein-Hawking entropy by $S_{BH}$, which scales as the horizon surface area $A$ of the black hole (entropy-area law).
By plugging into Eq.~\eqref{KanP}, we obtain
\be
\label{KenBH}
S_\kappa=\frac{1}{\kappa}\sinh\left(\kappa\hspace{0.4mm}S_{BH}\right).
\ee
Two comments are in order here: first, we notice that $S_\kappa$ as written in Eq.~\eqref{KenBH} is apparently symmetric for $\kappa\rightarrow-\kappa$. Thus, one can simply restrict to positive values of the entropic exponent. Furthermore, considering that deviations from the classical entropy are expected to be small (i.e. $\kappa\ll1$), it seems reasonable to expand $S_{\kappa}$ to the leading order as
\be
\label{app}
S_{\kappa}=S_{BH}+\frac{\kappa^2}{6}S_{BH}^3+\mathcal{O}(\kappa^4)\,,
\ee
where the zero-th order returns the entropy-area law, as expected.
The above approximation is useful to extract analytical solutions from 
Kaniadakis entropy-based equations, especially in the cosmological framework~\cite{Lymperis:2021qty}. We shall rely on Eq.~\eqref{app} for our next considerations, verifying a posteriori the validity of the condition $\kappa\ll1$. 

\subsection{Modified Friedmann equations}
Modified cosmology through Kaniadakis entropy~\eqref{KenBH} has been explored in~\cite{Lymperis:2021qty}. This study has revealed that new cosmological scenarios emerge based on corrected Friedmann equations, which contain extra terms that represent an effective dark energy sector depending on the model parameter $\kappa$. It was argued that the effective dark energy equation of state parameter deviates from standard $\Lambda$CDM cosmology at small
redshifts, and remains in the phantom regime during the history of
the Universe \cite{Lymperis:2021qty}. While achieving the same formal result, a geometric re-interpretation of these corrections has been provided in~\cite{Sheykhi:2023aqa}, motivated by the observation that entropy is a geometrical quantity and any
modification to it should change the geometry part of the field
equations \cite{Sheykhi:2023aqa}. In what follows, we stick to the latter derivation and consider a homogeneous and isotropic FRW flat geometry with metric\footnote{For the general case of a curved geometry, see~\cite{Lymperis:2021qty,Sheykhi:2023aqa}.} 
\be
\label{FRW}
ds^2=h_{\mu\nu}dx^\mu dx^\nu+\tilde r^2\left(d\theta^2+\sin^2\theta\, d\phi^2\right),
\ee
where $\tilde r=a(t)\hspace{0.2mm}r$, $x^0=t$, $x^1=r$, $h_{\mu\nu}=\mathrm{diag}\left(-1,a^2\right)$ and $a(t)$ is the time-dependent scale factor. 

Conceiving the Universe as a spherical thermodynamic system, 
the radius of the apparent horizon is $\tilde r_A=1/H=a/\dot a$, where
$H$ is the Hubble rate and the overdot denotes ordinary derivative with respect to the cosmic time $t$. In turn, the associated temperature follows from the definition of the surface gravity~\cite{Akbar:2006kj}
\be
K=-\frac{1}{\tilde r_A}\left(1-\frac{\dot{\tilde r}_A}{2H\tilde r_A}\right),
\ee
which gives
\be
\label{temp}
T_h=\frac{K}{2\pi}=-\frac{1}{2\pi\tilde r_A}\left(1-\frac{\dot{\tilde r}_A}{2H\tilde r_A}\right).
\ee

We now assume the Universe to be filled with matter and energy in the form of perfect fluid. Denoting the equilibrium energy density and pressure by $\rho$ and $p$, respectively, the energy-momentum tensor is
\be
T_{\mu\nu}=\left(\rho+p\right)u_\mu u_{\nu}+p\,g_{\mu\nu}\,,
\ee
where $u_\mu$ is the four-velocity of the fluid. The conservation equation for the total matter and energy content reads $\nabla_{\mu}T^{\mu\nu}=0$, which gives 
\be
\label{ce}
\dot \rho=-3H\left(\rho+p\right)
\ee
for the background~\eqref{FRW}. 

In order to derive the cosmological equations, let us employ the gravity-thermodynamic conjecture. Toward this end, we apply the first law of thermodynamics 
\be
\label{flt}
dE=T_hdS_h+WdV\,,
\ee
on the apparent horizon of entropy $S_h$, where $E=\rho V$ is the total energy in the spherical volume $V$ enclosed by the horizon. Due to the cosmic expansion, the work density done by a change
in the horizon radius is
\be
W=-\frac{1}{2}T^{\mu\nu}h_{\mu\nu}=\frac{1}{2}\left(\rho-p\right)\,.
\ee

We omit standard textbook calculations. Assuming that the entropy of the apparent horizon is in the form of Kaniadakis entropy~\eqref{app} and replacing the horizon radius of the black hole with the radius of
the apparent horizon, after some algebra we get from Eq.~\eqref{flt}~\cite{Sheykhi:2023aqa} 
\be
\label{FFE}
H^2-\kappa^2\hspace{0.2mm}\frac{\pi^2}{2\left(GH\right)^2}\simeq\frac{8\pi G}{3}\hspace{0.2mm}\rho\,,
\ee
where we have neglected the irrelevant contribution from the cosmological constant. This is the first Friedmann equation in Kaniadakis entropy-based cosmology \cite{Sheykhi:2023aqa}. 

Similarly, one can derive the second Friedmann equation by taking the time derivative of Eq.~\eqref{FFE} and using the continuity equation~\eqref{ce}. We get
\be
\label{SFE}
\dot H\left[1+\kappa^2\hspace{0.2mm}\frac{\pi^2}{2\left(GH^2\right)^2}\right]\simeq-4\pi G\left(\rho+p\right)\,.
\ee

It is worth noting that Eqs.~\eqref{FFE} and~\eqref{SFE} coincide with the leading order of the exact equations found in~\cite{Lymperis:2021qty}. Clearly, the Friedmann equations in standard cosmology are recovered for $\kappa\rightarrow0$. 

Before moving on to the study of the implications of Eqs.~\eqref{FFE} and~\eqref{SFE}, we remark that modified cosmic scenarios have also been investigated in Tsallis~\cite{TsallisCosm1,TsallisCosm2,TsallisCosm3,TsallisCosm4} and Barrow~\cite{BarrowCosm1,BarrowCosm2,BarrowCosm3,BarrowCosm4} entropies, inspired by non-extensive and quantum gravitational considerations, respectively. Here, we stress that 
Kaniadakis model has a relativistic foundation that makes it conceptually different from the other modified cosmologies. 

\section{Slow-roll inflation in Kaniadakis modification of Friedmann cosmology}
\label{Inf}

We now study inflation in Kaniadakis modification of Friedmann cosmology. Following the analysis of~\cite{Keskin,LucBarInf}, we consider the high-energy era of the Universe from the slow-roll condition perspective~\cite{OdSR} and assume the evolution to be driven by a scalar field $\phi$ with potential $V(\phi)$. In this setting, the characteristic parameters measuring inflation are the tensor-to-scalar ratio and the scalar spectral index of the primordial curvature perturbations, which for a minimal coupling are defined by~\cite{Inf1,Inf2}
\begin{eqnarray}
\label{r}
r&=&16\epsilon\,,\\[2mm]
n_s&=& 1-6\epsilon+2\eta\,,
\label{ns}
\end{eqnarray}
respectively. Here, we have denoted the {exact inflationary parameters by}
\begin{eqnarray}
\label{slow1}
\epsilon &=& - \frac{\dot H}{H^2}\,,\\[2mm]
\eta&=&-\frac{\ddot H}{2H\dot H}\,.
\label{slow2}
\end{eqnarray}

Under the canonical scalar field assumption, we can write the model lagrangian as 
\be
\mathcal L=X-V(\phi)\,,
\ee 
where 
\be
X=-\frac{1}{2}h^{\mu\nu}\partial_\mu\phi\hspace{0.2mm}\partial_\nu\phi\,,
\ee
and $V(\phi)$ are the kinetic and (spatially homogenous) potential terms, respectively. The energy density and pressure of the associated matter content inside the early Universe are now
\begin{eqnarray}
\label{def1}
\rho_\phi&=&\frac{\dot\phi^2}{2}+V(\phi)\,,\\[2mm]
\label{def2}
p_\phi&=&\frac{\dot\phi^2}{2}-V(\phi)\,.
\end{eqnarray}
They obey the continuity equation~\eqref{ce}, which can be recast in the Klein-Gordon-like form
\be
\label{KG}
\ddot \phi + 3H \dot\phi + \partial_\phi V(\phi) =0\,,
\ee
where we have used the shorthand notation $\partial_\phi V\equiv\frac{dV}{d\phi}$.

Assuming that the potential energy dominates all other forms
of energy, the slow-roll approximation is expressed as~\cite{Inf1,duality}
\be
\label{SRC}
\ddot \phi \ll H \dot \phi\,,\qquad\, \frac{\dot\phi^2}{2} \ll V(\phi)\,.
\ee
If we solve the first modified Friedmann equation~\eqref{FFE} with respect to $H$ and use the second of the above conditions, we obtain to the leading order in $\kappa$
\be
\label{Happ}
H\simeq\sqrt{{\frac{8\pi G}{3}\hspace{0.2mm}V}}+ \sqrt{\frac{27\pi}{2\hspace{0.2mm}G^{7}\hspace{0.2mm}V^{3}}}\frac{\kappa^2}{64}\,,
\ee
where we have omitted the field dependence of $V$ to streamline the notation. 

On the other hand, the second Friedmann equation~\eqref{SFE} along with Eqs.~\eqref{def1} and~\eqref{def2} give
\be
\label{Hdotde}
\dot H\simeq\left(-4\pi G+\frac{9\pi\kappa^2}{32G^3V^2}\right)\dot\phi^2\,. 
\ee
Hence, the parameters~\eqref{slow1} and~\eqref{slow2} take the form\footnote{{One might be tempted to naively neglect the $\dot{\phi}^2$ and $\ddot\phi$-terms in Eqs.~\eqref{eps} and~\eqref{eta} in the slow-roll approximation. However, the correct implementation of Eq.~\eqref{SRC} allows for such terms to be omitted only when compared to $V(\phi)$ and $H \dot{\phi}$, respectively. In fact, if one brutally sets $\dot{\phi}^2$ to zero in Eqs.~\eqref{Hdotde} and~\eqref{eps}, then the time derivative of the Hubble rate and the parameter $\epsilon$ would be identically vanishing. In turn, the prediction of the standard model of cosmology could not be retrieved, not even in the limit $\kappa \rightarrow 0$, where Kaniadakis modified equations reduce to the standard Friedmann equations. 
On the other hand, in our analysis we consistently use the first of the two slow-roll conditions in Eq.~\eqref{SRC} when solving the field dynamics Eq.~\eqref{KG}, while the second condition is taken into account to simplify the energy density and pressure, Eqs.~\eqref{def1} and~\eqref{def2}, respectively.
}}
\begin{eqnarray}
\label{eps}
\hspace{-2mm}\epsilon\hspace{-0.5mm}&\simeq&\hspace{-0.5mm}\left(\frac{3}{2V}-\frac{27\kappa^2}{128}\frac{1}{G^4V^3}\right)\dot\phi^2\,,\\[2mm]
\nonumber
\label{eta}
\hspace{-2mm}\eta\hspace{-0.5mm}&\simeq&\hspace{-0.5mm}\left[-\sqrt{\frac{3}{8\pi G\hspace{0.2mm}V}}\,\ddot\phi\right.\\[2mm]
&&+\left.\sqrt\frac{3}{2\pi\hspace{0.2mm}G^9\,V^7}\frac{9\kappa^2}{512}\left(\ddot\phi\,V-2\dot\phi^2\,\partial_\phi V\right)\right]\frac{1}{\dot\phi}\,\,. 
\end{eqnarray}
As observed in~\cite{Keskin}, these two parameters are to be computed at the horizon crossing, where the fluctuations of the inflaton freeze. 

The amount of inflation that occurs is measured by
the number $N$ of e-folds~\cite{Carrol}
\be
\label{N}
N=\int_{t_i}^{t_f} H(t) dt\,,
\ee
where we have denoted the initial (final) time of the inflation by
$t_i$ $(t_f)$. Since the primordial fluctuations of the inflaton 
are relevant during the horizon crossing and the power spectrum of the inflation dynamics is evaluated at this point, the beginning time of the inflation is set to the horizon crossing time, i.e. $t_i\equiv t_c$. In terms of the field, this amounts to saying $\phi_i\equiv\phi_c$, which allows us to write Eq.~\eqref{N} in the equivalent form
\be
\label{efold}
N=\int_{\phi_c}^{\phi_f} \frac{H}{\dot\phi} \,d\phi\,, 
\ee 
where we have defined $\phi_c\equiv \phi (t_c)$ and 
$\phi_f\equiv \phi (t_f)$. 

The dynamics of the inflaton field is governed by the potential $V(\phi)$. In our model, we assume a power-law potential in the form
\be
\label{pot}
V=V_0 \phi^n\,,
\ee
where $V_0>0$ is a constant with dimension $[E]^{4-n}$ associated with the energy scale of inflation $E_{inf}\simeq \mathcal{O}(10^{15})\,\mathrm{GeV}$ and $n>0$ is the power-term. From the latest data, we know that inflationary models with $n\sim\mathcal{O}(10^{-1}\div 1)$ are observationally favored, while $n\ge2$ tends to be excluded in the minimal coupling setting. In the following, we assume $n\simeq1$. 

{Before moving on, it is important to clarify that
two different perspectives are typically taken into account in the study of the inflationary scenario. On one hand, one can reconstruct the potential $V$ by requiring the Universe to evolve according to a specific dynamics, which is set by a fixed time-dependence of the Hubble rate (or, equivalently, of the scale factor $a$). In this approach, which is usually referred to as \emph{dynamically motivated approach}, inflationary models are classified based on the corresponding Universe evolution.  An example of this approach appears in~\cite{duality} by considering the scale factor as $a(t)=a_0t^{m}$ for a specific choice of the power term $m$.
However, as noted in~\cite{duality}, there are two main issues with this model. First, the scalar-to-tensor ratio is found to be larger
than the limits set by the BICEP and Planck data. Furthermore, the parameters for the slow-roll are constant.} 

{On the other hand, there is an equally valid method, in which one firstly sets the scalar field's potential and then solve the Friedmann equations accordingly. This type of method is known as the \emph{potential motivated approach} and is largely adopted in literature on inflation  
in extended gravity, where deviations of the dynamics of the scale factor from the typical GR predictions are contemplated~\cite{Keskin,SA1,SA2,SA3,LucBarInf,N2}.
In this case, inflationary models are categorized according to the form of the scalar field potential, which is fixed a priori (see also~\cite{duality} for more details on the issue). 
By setting Eq.~\eqref{pot}, we are here considering an example of the second approach. And, in fact, the potential $V$ enters the model dynamics through the energy density and pressure of the scalar field, Eqs.~\eqref{def1} and~\eqref{def2}, respectively, which must be included in the amount of total energy density and pressure to consider in the right side of the modified Friedmann equations~\eqref{FFE} and~\eqref{SFE}.}

Analytical solutions of the inflationary observable indices can be extracted by expressing $\dot\phi$ and $\ddot \phi$ as functions of the scalar field $\phi$.  Toward this end, we use the first of the two slow-roll conditions in Eq.~\eqref{SRC} to simplify the dynamics~\eqref{KG} as
\be
\dot\phi\simeq-\frac{1}{3H}\hspace{0.2mm}\partial_\phi V\,. 
\ee
By substitution of Eqs.~\eqref{Happ} and~\eqref{pot}, we are led to
\be
\dot\phi\simeq-\frac{n}{2}\sqrt{\frac{V_0}{6\pi G}}\phi^{(n-2)/2}
+\frac{3\hspace{0.2mm}n\kappa^2}{512}\sqrt{\frac{3}{2\pi\left(G^9\hspace{0.3mm}V_0^3\right)}}\phi^{-(2+3n)/2}
\ee

Let us impose that inflation (and counting of e-folds $N$) ends when 
$\epsilon(\phi_f)\sim1$. We first recast Eq.~\eqref{eps} as
\be
\epsilon(\phi_f)\simeq\frac{1}{16\pi \hspace{0.2mm}G\hspace{0.2mm}\phi_f^2}-\frac{27\kappa^2}{2048\pi\hspace{0.2mm}G^5(V_0\phi_f^2)^2}\,,
\ee
where we have used the power-law potential~\eqref{pot} with $n\simeq1$. Solving with respect to $\phi_f$, we obtain
\be
\phi_f\simeq\frac{1}{4\sqrt{\pi G}}-\frac{27\kappa^2}{64V_0^2}\sqrt{\frac{\pi}{G^7}}\,,
\ee 
where we have considered the only solution that recovers the correct limit for $\kappa\rightarrow0$. 

The value $\phi_c$ of the inflaton at the horizon crossing can be then estimated by Eq.~\eqref{efold}, which gives
\be
\label{ansatz}
N\simeq -\frac{1}{4}+4\pi \hspace{0.2mm}G \phi_c^2+\frac{9\pi \kappa^2\left[3+\log\left(16\pi \hspace{0.2mm}G\phi_c^2\right)\right]}{32\hspace{0.2mm}G^3V_0^2}\,.
\ee

To solve this equation, we make the ansatz that the two terms in the  
square brackets are of the same order. We check a posteriori the validity of this assumption (see below Eq.~\eqref{nsb}). Under the above condition, we obtain
\be
\label{phic}
\phi_c\simeq\frac{1}{4}\sqrt{\frac{1+4N}{\pi\hspace{0.2mm}G}}-\frac{1.2\kappa^2}{V_0^2}\sqrt{\frac{\pi}{\left(1+4N\right)G^7}}\,.
\ee
By use of this relation, we can cast the tensor-to-scalar ratio~\eqref{r} and the scalar spectral index~\eqref{ns} the in terms of the e-folding time $N$ (see~\cite{Liddle:2003as} for observational discussions on the limit of e-folds). To this aim, we observe that
\begin{eqnarray}
\epsilon(\phi_c)&\simeq&\frac{1}{1+4N}+\frac{6.2\pi \kappa^2}{G^3\left[V_0\left(1+4N\right)\right]^2}\,,\\[2mm]
\eta(\phi_c)&\simeq&\frac{1}{-1-4N}-\frac{4\pi\kappa^2}{G^3\left[V_0\left(1+4N\right)\right]^2}\,.
\end{eqnarray}
which can be inserted into Eqs.~\eqref{r} and~\eqref{ns} to give
\begin{eqnarray}
\label{rbis}
r_\kappa&\simeq&\frac{16}{1+4N}+\frac{10^2\pi\kappa^2}{G^3\left[V_0\left(1+4N\right)\right]^2}\,,\\[2mm]
\label{nsbis}
n_{s_\kappa} &\simeq&\frac{4N-7}{1+4N}-\frac{45.5\pi \kappa^2}{G^3\left[V_0\left(1+4N\right)\right]^2}\,.
\end{eqnarray}

We now notice that the above description of inflation in Kaniadakis modification of Friedmann cosmology is phenomenologically consistent, provided that Eqs.~\eqref{rbis} and~\eqref{nsbis} match with the {latest BICEP and Planck observations, which set the experimental bounds~\cite{ConPlanck}
\be
r<0.032 \hspace{0.8cm} (95\%\,\, \mathrm{CL})\,, 
\label{rb}
\ee
and~\cite{ExpData}
\be
\label{nsb}
n_s=0.965\pm 0.004 \,\,\,\,\,\,\, (68\%\,\, \mathrm{CL})\,. 
\ee}

\begin{figure}[t]
\centering
\hspace{-4mm}\includegraphics[width=8.6cm]{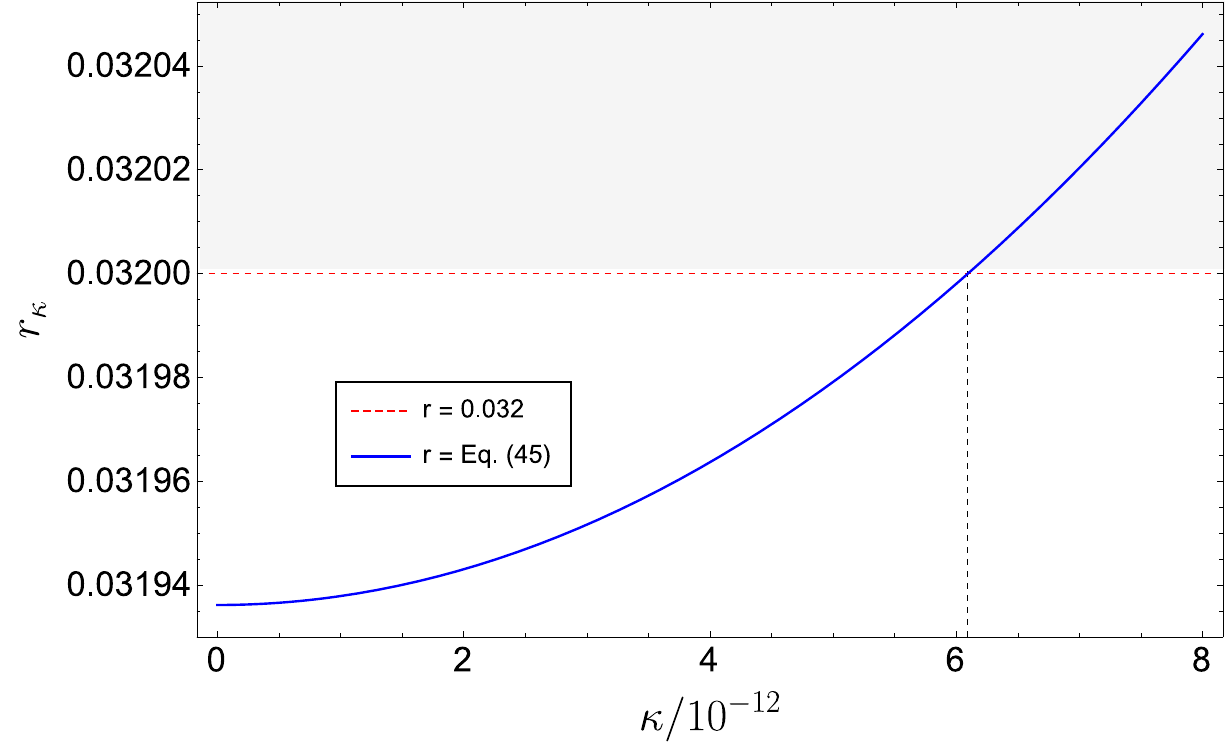}
\caption{Plot of $r$ versus the (rescaled) Kaniadakis parameter. The grey-shaded region is excluded by the observational constraint~\eqref{rb}, which is represented by the horizontal red-dashed line. The ensuing upper bound $\kappa\lesssim6.09\times10^{-12}$ on Kaniadakis parameter is indicated by the vertical black-dashed line. We set $N$ of the same order as the highest value in~\cite{Liddle:2003as}. }
\label{Fig1}
\end{figure}

{The behavior of $r_\kappa$ in Eq.~\eqref{rbis} is plotted 
versus the (rescaled) Kaniadakis parameter in Fig.~\ref{Fig1}.  We see that the $\kappa$-corrected tensor-to-scalar ratio is observationally consistent as far as $\kappa\lesssim6.1\times10^{-12}$. Moreover, this constraint is in agreement with the prediction of the scalar spectral index too. Indeed, by requiring that $n_{s_\kappa}$ in Eq.~\eqref{nsbis} be equal to the experimental value~\eqref{nsb}, we find $\kappa\simeq5.8\times10^{-12}$, which is consistent with the above limitation. }

{It remains to be seen whether the ansatz below Eq.~\eqref{ansatz} is valid. Toward this end, by using Eq.~\eqref{phic}, we get $\log\left(16\pi\hspace{0.2mm}G\phi_c^2\right)/3\sim\mathcal{O}(1)$, which confirms our assumption and the physical consistency of our analysis. Therefore, we conclude that Kaniadakis modification of Friedmann cosmology allows for the slow-roll inflation phase in a Universe driven by a scalar field with power-law potential ($n\sim\mathcal{O}(1)$), setting the constraint 
\be
\label{Constraint}
\kappa\lesssim\mathcal{O}(10^{-12}\div10^{-11})
\ee
on the entropic parameter. More discussion on this result compared to the literature can be found in the concluding section. }

\section{Growth of Cosmological Perturbations}
\label{Growth}
In this section we explore the influence of the modified Kaniadakis entropy on the growth of cosmological perturbations and power spectrum at the early stages of the Universe. Following~\cite{Gr3}, we now consider, for simplicity, a universe filled with pressure-less matter. This assumption does not affect the generality of our next considerations. For later convenience, we recast the modified Friedmann equations~\eqref{SFE} as
\be
\label{addot}
\frac{\ddot a}{a}\simeq-\frac{4\pi G}{3}\hspace{0.2mm}\rho+\frac{15\pi\kappa^2}{32G^3\rho}\,,
\ee
where we have used
\be
\frac{\ddot a}{a}=\dot H+H^2\,,
\ee
along with the first Friedmann equation~\eqref{FFE}. The conservation equation~\eqref{ce} for the dust matter takes the form
\be
\label{cebis}
\dot\rho+3H\rho=0\,.
\ee

\subsection{Density contrast}

To study the evolution of the cosmological perturbations, we use the the Top-Hat Spherical Collapse (SC) model~\cite{SCM}.
In this formalism, one considers 
a uniform and spherical symmetric perturbation in an expanding
background and analyzes the 
growth of perturbations
in a spherical region of radius $a_p$ and density $\rho^c$ by resorting to the same Friedmann equations as for the underlying theory of gravity~\cite{Gr1,Gr2,Gr3}. At time $t$, the density of the fluid in this region can be written as~\cite{Gr3}
\be
\rho^c=\rho(t)+\delta \rho\,,
\ee
where $\delta\rho$ denotes the density fluctuation. The conservation equation reads
\be
\label{ceter}
\dot\rho^c+3h\rho^c=0\,,
\ee
where $h=\dot a_p/a_p$ is the local Hubble rate of the 
perturbed region. Additionally, 
since Eq.~\eqref{addot} is valid in the whole spacetime, it can be specifically written for the perturbed region to give
\be
\label{addotp}
\frac{\ddot a_p}{a_p}\simeq-\frac{4\pi G}{3}\hspace{0.2mm}\rho^c+\frac{15\pi\kappa^2}{32G^3\rho^c}\,.
\ee

Let us now define the density contrast $\delta$ of the fluid in the Universe by
\be
\label{deltadef}
\delta=\frac{\rho^c}{\rho}-1=\frac{\delta\rho}{\rho}\ll1\,,
\ee
where we have reasonably assumed the fluctuation to be much smaller than the density itself (linear regime). 
Deriving the above relation with respect to $t$ and using Eqs.~\eqref{cebis} and~\eqref{ceter}, we obtain
\be
\label{ddelta}
\dot\delta=3\left(1+\delta\right)(H- h)\,,
\ee
which by further derivation gives
\be
\label{dddelta}
\ddot\delta=3\left(1+\delta\right)(\dot H-\dot h)+\frac{\dot\delta^2}{1+\delta}\,.
\ee
This is the differential equation for the evolution of the matter
perturbations. 

We can observe that the simultaneous usage of Eqs.~\eqref{addot},~\eqref{addotp} and~\eqref{deltadef} leads to
\be
\dot H-\dot h\simeq h^2-H^2+\left(\frac{4\pi G}{3}\hspace{0.2mm}\rho+\frac{15\pi\kappa^2}{32G^3\rho}\right)\delta\,,
\ee
where we have expanded to the leading order in $\delta$ due to the linear regime we are working in. Hence, we are allowed to recast the differential equation~\eqref{dddelta} as
\be
\label{difeq}
\ddot\delta+2H\dot\delta-\left(4\pi G\rho+\frac{45\pi\kappa^2}{32G^3\rho}\right)\delta=0\,.
\ee

In order to analyze the impact of Kaniadakis entropy on the evolution of the density contrast, it is convenient to express Eq.~\eqref{difeq} in terms of the redshift parameter 
\be
z=\frac{1-a}{a}\,,
\ee 
where we have set the present value of the scale factor to unity. 
First, we recall that the matter energy density is given by 
\be
\rho=\rho_0\left(1+z\right)^3\,.
\ee 
Then, we replace the time derivatives with derivatives respect to $a$ (denoted by the prime). Toward this end, we note that
\begin{eqnarray}
\dot\delta&=& a H \delta'\,,\\[2mm]
\ddot\delta&=&a^2H^2
\delta''
+a\left(H^2-4\pi G\rho+\frac{2\pi^3\kappa^2\rho}{G H^4}\right)\delta'\,.
\end{eqnarray}
Inserting into Eq.~\eqref{difeq}, we are finally led to
\be
\delta''+\frac{3}{2a}\left(1+\frac{9\kappa^2}{64 G^4\rho^2}\right)\delta'-\frac{3}{2a^2}\left(1+\frac{9\kappa^2}{32 G^4\rho^2}\right)\delta=0\,.
\ee
It can be easily checked that, for $\kappa\rightarrow0$, this equation reduces to
\be
\delta''+\frac{3}{2a}\delta'-\frac{3}{2a^2}\delta=0\,,
\ee
which correctly reproduces the result of the SMC in the absence of the cosmological constant~\cite{SCM}.

\begin{figure}[t]
\centering
\hspace{-4mm}\includegraphics[width=8.6cm]{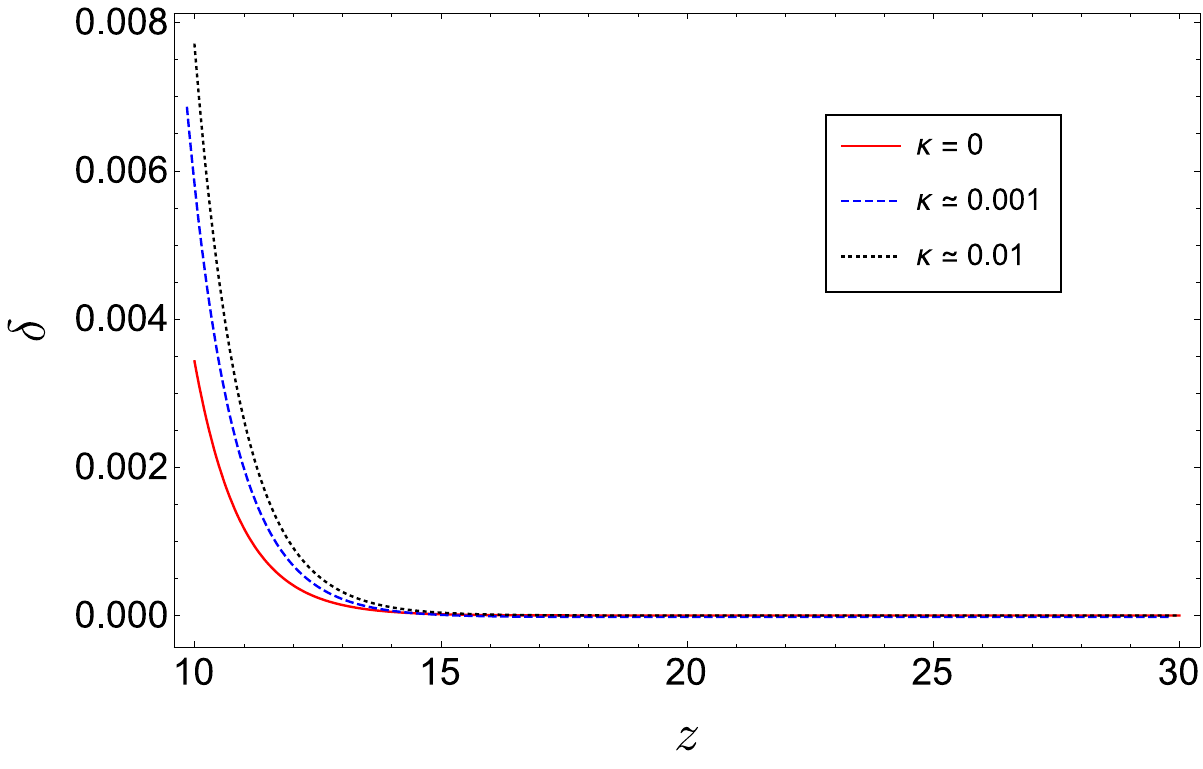}
\caption{Plot of the density contrast $\delta$ versus $z$. The values of $\kappa$ have been overestimated to graphically appreciate Kaniadakis corrections. We have assumed the same initial conditions as in~\cite{Gr3} and worked in units of $G$.}
\label{Fig3}
\end{figure}

The behavior of the matter density contrast $\delta$ versus  redshift is plotted in Fig.~\ref{Fig3} for different values of $\kappa$ and in comparison with the prediction of the SMC ($\kappa=0$).  We can see that Kaniadakis corrections influence the growth of cosmological perturbations (in particular at lower redshift), in such a way that the higher $\kappa$, the faster the growth rate. 
Such a result can be understood by observing that, according to Kaniadakis' prescription~\eqref{KenBH}, the 
modified entropy (and, thus, the
number of holographic degrees of freedom) of the Universe gets increased compared to the standard Boltzmann-Gibbs scenario, supporting a more rapid growth of fluctuations in energy density. 

It is interesting to note that a similar behavior has been found in~\cite{Gr3} in the framework of quantum gravity deformations of the entropy-area law, motivated by the intricate fractal structure of the apparent horizon in Barrow cosmology. On the other hand, corrections induced by non-extensive Tsallis statistics may result in a faster or slower growth of perturbations, depending on the value of the model parameter~\cite{Gr3}. 

\subsection{Power spectrum}
To stress test the phenomenological consistency of our model, let us now consider the impact of Kaniadakis entropy on the primordial power spectrum $P_\zeta$ for scalar perturbations. This is defined by the 2-point correlator of scalar perturbations as
\be
\langle\zeta^2(\vec{r}) \rangle = \int dk\, \frac{P_\zeta(k)}{k}\,,
\ee
where the energy scale $k$ can be expressed in terms of e-folds by
$N(k)=\log\left(k_{final}/k\right)$~\cite{Arb}.

The CMB power spectrum can be described by the Harrison-Zeldovich (HZ) fit~\cite{HZ}
\be
\label{HZ}
P_{\zeta}^{HZ}(k) = A_s\left(\frac{k}{k_*}\right)^{n_s-1}\,,
\ee
which is parametrized by the power amplitude $A_s\simeq10^{-9}$ at the pivot scale $k_*=0.05\,\mathrm{Mpc}^{-1}$, or in the slow-roll approximation by
\be
\label{SR}
P_{\zeta}^{SR}\simeq P_0 \log^2\left(\frac{k}{k_{final}}\right)= P_0 N^2,
\ee
where $P_0=\emph{const.}$

In the slow-roll approximation, the power spectrum $P_\zeta(N)$
is connected to the potential $V$ via the standard relation~\cite{PV}
\be
P_\zeta(N)=\frac{V^2}{12\pi^2 M_p^4 }\left(\frac{dV}{dN}\right)^{-1}\,.
\ee

In the case of the potential~\eqref{pot} with $\phi$ evaluated at crossing, one can show by straightforward computations that $P_\zeta(N)$ is consistent with $P_{\zeta}^{SR}(N)$ as concerns the quadratic scaling in $N$ of the zero-th order term in $\kappa$, provided that the power-term $n=2$. However, as discussed below Eq.~\eqref{pot},  inflation scenarios with $n\ge2$ are statistically disfavored. 

It must be observed that the simple form~\eqref{HZ} (or~\eqref{SR}) 
of the primordial power spectrum yields a reasonably good fit of experimental data with a remarkably small number of parameters. Nevertheless, as emphasized in~\cite{deBlas}, it does not provide the best observational fit, which is instead obtained by allowing the primordial power spectrum to deviate from the simple power-law form with a running of the spectral index. This happens because of the possibility of suppressing the primordial power spectrum at large
scales. 

A non-trivial improvement of Eq.~\eqref{HZ} can be gained by considering a multi-parametric broken-power-law (bpl) spectrum in the form\footnote{For completeness, we mention that an alternative form considered by Planck collaboration for the primordial power spectrum is given by a simple power-law multiplied by an exponential cut-off~\cite{deBlas}. This functional form is typically adopted in scenarios where the slow roll follows an stage of kinetic
energy domination, see e.g.~\cite{eg}.}~\cite{deBlas}
\be
\label{mpmod}
P^{bpl}_\zeta=
        A_{low}\left(\frac{k}{k_*}\right)^{n_s^{bpl}-1+\delta}\,, 
  \ee
for $k$ less than or equal to a certain cutoff $k_b$, 
while the spectrum~\eqref{HZ} is kept unchanged for $k$ higher than this scale. Here, we have defined $A_{low}=A_s\left(k_b/k_*\right)^{-\delta}$, so that $P^{bpl}_\zeta$ is continuous at $k=k_b$. 
For this model, the best fit to the TT+lowP Planck
data is obtained for $n_s^{bpl}=0.9658$, $\log\left(k_b/\mathrm{Mpc}^{-1}\right)=-7.55$ and $\delta=1.14$~\cite{deBlas}. 

Interestingly enough, our model based on Kaniadakis modified Friedmann equations naturally incorporates the feature of a running spectral index, as it appears from Eq.~\eqref{nsbis} with a time-dependent $\kappa$ (see the discussion in the Conclusion). 
The correspondence with the best-fit data model
can be further elaborated by rewriting Eq.~\eqref{nsbis} in the equivalent form
\be
\label{eqform}
n_{s_\kappa}=n_s+ C_\kappa\,,
\ee
where $n_s$ is the standard spectral index for $\kappa=0$ and $C_\kappa$ the correction induced by Kaniadakis entropy. One can see that the latter term plays the same role as the $\delta$-shift in the bpl spectrum~\eqref{mpmod}. In other terms, the effects of the modified spectrum~\eqref{mpmod} can be reproduced in Kaniadakis language by retaining the HZ fit~\eqref{HZ} and 
expressing the spectral index in terms of its Kaniadakis counterpart. 
To see this, let us observe that, after
substitution of Eq.~\eqref{eqform} into~\eqref{HZ}, the exponent of the HZ fit takes the form
\be
n_s-1=n_{s_\kappa}-C_\kappa-1\,.
\ee
The right side gives the same effects as the $\delta$-shift in Eq.~\eqref{mpmod}, provided that 
\be
n_{s_\kappa}-C_\kappa-1=n_s^{bpl}-1+\delta\,.
\ee
Furthermore, consistency with Planck data requires that
$n_{s_\kappa}$ and $n_s^{bpl}$ be equal to the observational value~\eqref{nsb}, which yields 
\be
\delta=-C_\kappa=\frac{45.5\pi \kappa^2}{G^3\left[V_0\left(1+4N\right)\right]^2}\,.
\ee

The above equation can be solved for 
 $N, V_0$ fixed as in Sec.~\ref{Inf} and $\delta=1.14$~\cite{deBlas} to give $\kappa\simeq5.9\times10^{-12}$, which is in line with the bound displayed below Eq.~\eqref{nsb}. We conclude that 
our model provides an alternative characterization of the 
bpl spectrum~\eqref{mpmod}, which is still compatible with Planck observations (see also the next Section for more discussion).

\section{Conclusion and discussion}
\label{DC}

Kaniadakis entropy provides a coherent and self-consistent attempt 
to generalize Boltzmann-Gibbs-Shannon statistics to the relativistic framework. Furthermore, its holographic usage in cosmology underlies the derivation of the modified Friedmann equations, which predict an interesting phenomenology. In this work we have studied 
implications of Kaniadakis modification of Friedmann cosmology on the slow-roll inflationary era. We have assumed the energy content of the Universe to be represented by a scalar field with a perfect fluid form and a power-law potential. Phenomenological consistency of our model has been explored by computing the scalar spectral index and tensor-to-scalar ratio at the horizon crossing. From the latest BICEP and Planck data, we have obtained $\kappa\lesssim\mathcal{O}(10^{-12}\div10^{-11})$, to be
compared with other estimates from Baryon Acoustic Oscillation (BAO)~\cite{Hernandez-Almada:2021aiw}, Type Ia supernova\footnote{It is worth noting that 
the values of $\kappa$ given in~\cite{Hernandez-Almada:2021aiw,CosmKan4} are in terms of the re-scaled parameter $\beta=\kappa\frac{M_p^2}{H_0^2}$, where $H_0$ is the present Hubble rate.}~\cite{Hernandez-Almada:2021aiw} and baryogenesis~\cite{KanLuc} measurements, among others (see Tab~\ref{Tab1}). 

\begin{table}[t]
  \centering
    \begin{tabular}{|c| c c|}
    \hline
\, $|\kappa|$\,\, &\, Physical framework \, &\,  Ref.\, \\
        \hline
        \,$10^{-125}$\, & 
             Baryon Acoustic Oscillations (BAO) & \cite{Hernandez-Almada:2021aiw} \\[1.5mm]
           \hline
           \,$10^{-125}$\, & 
             CC+SNIa+BAO & \cite{Hernandez-Almada:2021aiw} \\[1.5mm]
           \hline
           \,$10^{-124}$\, & Cosmological constant (CC) & \cite{Hernandez-Almada:2021aiw} \\[1.5mm]
           \hline
            \,$10^{-124}$\, & 
            Type Ia supernova (SNIa) & \cite{Hernandez-Almada:2021aiw} \\[1.5mm]
            \hline
            \,$10^{-123}$\, & 
             Hubble data & \cite{CosmKan4} \\[1.5mm]
           \hline
              \,$10^{-123}$\, & 
             Strong lensing systems & \cite{CosmKan4} \\[1.5mm]
           \hline
              \,$10^{-123}$\, & 
             HII galaxies & \cite{CosmKan4} \\[1.5mm]
           \hline
            \,$10^{-83}$\, & Baryogenesis
              & \cite{KanLuc} \\[1.5mm]
              \hline
    \end{tabular}
  \caption{Cosmological constraints on Kaniadakis parameter.}
  \label{Tab1}
\end{table}
Although all these constraints on $\kappa$ lie within the domain fixed by the upper bound~\eqref{Constraint}, they span
a relatively wide range.  Despite not being contemplated in the original $\kappa$-statistics, this gap can be understood by allowing Kaniadakis parameter to have a running behavior, namely to vary with the energy (or time) scale. Such a behavior is actually what one expects from the holographic application of the relativistic Kaniadakis' prescription. In fact, entropy quantifies the physical degrees of freedom of a system -- the Universe in our cosmological setup. In the standard model of cosmology, it is typically assumed that the dynamics of the early Universe was initially set by relativistic energy content (radiation-dominated era), which was later exceeded by the energy density of the non-relativistic matter (matter-dominated era) as the Universe cooled down. It is then natural to apply the same paradigm to the entropic degrees of freedom. In the ensuing scenario, 
the relativistic nature of entropy would be quantified by a decreasing function of time, $\kappa\equiv \kappa(t)$ (or, equivalently, by an increasing function of the energy scale, $\kappa\equiv\kappa(E)$), in a way that it is maximal ($\kappa\sim\mathcal{O}(1)$) in the earliest stages of the Universe's existence, while it recovers the standard Boltzmann-Gibbs-Shannon behavior ($\kappa\simeq0$) at present time. This conjecture would explain in a unified picture the predictions $\kappa\lesssim\mathcal{O}(10^{-12}\div10^{-11})$ inferred from inflation ($E_{inf}\simeq(10^{15}\div10^{16})\,\mathrm{GeV}$), $\kappa\simeq\mathcal{O}(10^{-83})$ from Baryogenesis~\cite{KanLuc} ($100\,\mathrm{GeV}\lesssim E_{bar}\lesssim10^{12}\,\mathrm{GeV}$~\cite{Baryo}) and $\kappa\simeq\mathcal{O}(10^{-125})$ from Baryon Acoustic Oscillations measurements~\cite{Hernandez-Almada:2021aiw} (pre-recombination stage, $E_{rec}\simeq(10^{-1}\div 1)\,\mathrm{eV}$). In passing, we mention that a similar analysis has been recently conducted within the framework on non-extensive Tsallis thermodynamics with varying exponent~\cite{Varexp,LucBlas1,LucBlas2}. In that case, the running behavior is motivated by quantum field theoretical considerations associated with renormalization group flow, which is in principle unavoidable when one attempts to incorporate Tsallis entropy into a general framework that would be consistent with quantum gravity. 
Clearly, to substantiate this view in our case, one should consider an ab initio investigation of Kaniadakis modification of Friedmann cosmology equipped with a running parameter. This goes beyond the scope of the present analysis and will be explored elsewhere. 

We have then explored the effects of the modified Kaniadakis entropic corrections on the growth of
perturbations in the early stages of the Universe. Employing the spherically symmetric collapse formalism and working in the linear regime, we have extracted
the differential equation for the evolution of the matter
perturbations. Furthermore, for the matter density contrast we have observed that the modified profile of the growth of perturbations differs from the standard cosmological model, with increasing values of $\kappa$ corresponding to a faster growth of perturbations. A similar outcome has been recently exhibited in~\cite{Pert1,Pert2,Pert3} in the context of both Tsallis and Barrow Cosmologies.

{To further explore the phenomenological consistency of our model, we have finally studied the effects of Kaniadakis entropy on the primordial power spectrum for scalar perturbations. We have found that our model is compatible with the broken-power-law (bpl) spectrum~\eqref{mpmod}, which provides the best (although not statistically favored) observational fit. Clearly, as discussed above, a full correspondence can only be established after incorporating a dynamical $\kappa$ into the theory.}

Further aspects are yet to be investigated. In~\cite{Imp} a new description of inflation has been proposed by merging the Starobinsky model, the Appleby-Battye-Starobinsky parameterization of dark energy and a correction arising from the modified $F(R)$ gravity. It would be interesting to compare our Eqs.~\eqref{rbis} and~\eqref{nsbis} with predictions of this model, also in light of the Kaniadakis holographic description of dark energy proposed in~\cite{Drepanou}. Moreover, it is suggestive to examine Kaniadakis influence on the growth of perturbations and structure formation in relation to results from other deformed cosmologies based on extended gravity~\cite{Gr3,MCo1,MCo2}. 
Finally, to better test our model of cosmic inflation, 
we are interested in
studying Kaniadakis corrections 
to primordial local non-Gaussianity. In~\cite{Bravo} (and references therein)
it has been pointed out that 
Maldacena's consistency relation $f_{NL}=5(1-n_s)/12$
between the amount of local squeezed non-Gaussianity $f_{NL}$ and the spectral index $n_s$~\cite{Malda} cannot be directly observed.  In fact, the quantity that is actually observable is
\be
\label{62}
f_{NL}^{obs}=0+\mathcal{O}\left(\frac{k_L}{k_S}\right)^2\,,
\ee
where $k_{L(S)}$ denote the long (short) wavelength modes of the inflaton in Fourier space and the $\mathcal{O}\left(\frac{k_L}{k_S}\right)^2$ terms arise from non-primordial phenomena, such as gravitational lensing and redshift perturbations. The reason why Eq.~\eqref{62} holds is that the primordial value of $f_{NL}$ predicted by Maldacena's consistency in the co-moving gauge is canceled out by a correction $-5(1-n_s)/12+\mathcal{O}\left(\frac{k_L}{k_S}\right)^2$, caused by a change of coordinates to render observables gauge invariant. Of course, the simplification of the two terms occurs, leading to Eq.~\eqref{62},  provided that the prediction of primordial non-Gaussianity corresponds to Maldacena's consistency relation. The latter is known to be true for \emph{attractor} models of single-field inflation, i.e. models in which 
every background quantity during inflation
is set by a single parameter (e.g. the Hubble rate),
regardless of the initial conditions. Therefore, any appreciable measurement of local non-Gaussianity would rule out attractor single-field models of slow-roll inflation. On the other hand, more exotic descriptions of inflation, such as multi-field or non-attractor models, do not satisfy Maldacena's condition~\cite{Nonat,Nonat2} and would potentially be compatible with observations of primordial local non-Gaussianity. In this scenario, it is desirable to understand whether our Kaniadakis model of inflation falls within the latter class and, in that case, how the constraint~\eqref{Constraint} reconciles with the current bound on local non-Gaussianity from BICEP and Planck~\cite{ConPlanck}. Work along these directions is under active consideration and is left for future investigations.

\acknowledgements
GGL acknowledges the Spanish ``Ministerio de Universidades''
for the awarded Maria Zambrano fellowship and funding received
from the European Union - NextGenerationEU.
He is also grateful for participation to LISA cosmology Working group.
GGL and GL acknowledge networking support from the COST Action CA18108 ``Quantum Gravity Phenomenology in the Multimessenger Approach''. 


\end{document}